\documentclass{iaus}
\usepackage{graphicx}

\title[Novae and accreting WDs] %% give here short title %%
{Novae and accreting white dwarfs \\ as progenitors of Type Ia supernovae}

\author[Mariko Kato]   %% give here short author list %%
{Mariko Kato}
\affiliation{Keio University, \\ 4-1-1, Hiyoshi, Kouhoku-ku
Yokohama, 223-8521 Japan \\ email: {\tt mariko@educ.cc.keio.ac.jp} \\[\affilskip]
}
\pubyear{2011}
\volume{281}  %% insert here IAU Symposium No.
\jname{Binary Paths to the Explosions of Type Ia Supernovae}
\editors{R. Di Stefano \& M. Orio, eds.}
\begin{document}

\maketitle

\begin{abstract}
I review various phenomena associated with mass-accreting white dwarfs (WDs) 
in relation to progenitors of type Ia supernovae (SNe Ia). 
The WD mass can be estimated from light curve analysis in multiwavelength 
bands based on the optically thick wind theory. 
In the single degenerate scenario of SNe Ia, two main channels are known, i.e.,
WD + main sequence (MS) channel and WD + red giant (RG) channel. 
In each channel, a typical binary undergoes three evolutional 
stages before explosion, i.e., the wind phase, supersoft X-ray source (SSS) phase, 
and recurrent nova phase in order of time because the accretion rate decreases with time 
as the companion mass decreases. 
We can specify some accreting WDs as the corresponding stage of evolution.
Intermittent supersoft X-ray source like RX\,J0513.9$-$6951 and 
V\, Sge are corresponding to the wind phase objects. 
For the SSS phase Cal 87-type objects correspond to the WD+MS channel.  
For the WD + RG channel, 
soft X-ray observations of early type galaxies gave a statistical   
evidence of SSS phase binaries.  
Recurrent novae of U Sco-type and RS Oph-type correspond to the 
WD + MS channel and WD + RG  channel, respectively. 
Majority of recurrent novae host a very massive WD ($\gtrsim 1.35~M_\odot$) and often 
show a plateau phase in optical light curve correspondingly to the long lasted 
supersoft X-ray phase: These  properties are indications of increasing WD masses. 
\end{abstract}

%%%%%%%%%%%%%%%%%%%%%%%%   Figure  1    %%%%%%%%%%%%%%%%%%%%%%%%%%%%%%
\begin{figure}[t]
% \vspace*{-2.0 cm}
\begin{center}
 \includegraphics[width=2.5in]{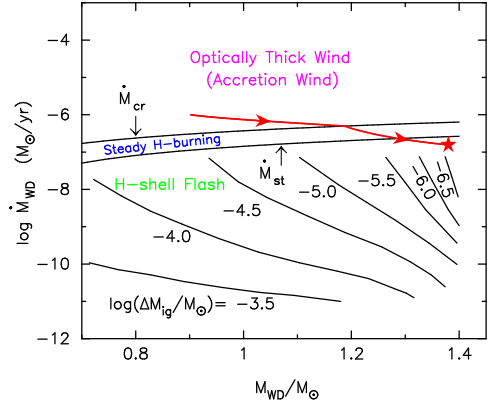}
% \vspace*{-1.0 cm}
 \caption{
A typical evolutionary path ({\it red solid} line) of an SN Ia progenitor
(in the SD scenario) on the map of response of WDs to mass accretion rate. 
Starting from the accretion wind phase, in which strong optically thick
winds blow from the WD, the binary enters the supersoft X-ray phase,
which is a narrow region between the accretion rate is $\dot M_{\rm cr} >
\dot M_{\rm WD} > \dot M_{\rm st}$.
When the mass accretion rate decreases, to less than the value to keep steady
hydrogen-burning ($\dot M_{\rm st}$), the binary enters the recurrent nova
(weak shell flashes) phase. The WD explodes at the star mark as an SN Ia 
(taken from \cite[Hachisu et al. 2010]{hac10kn}).
}
% \label{accmap_z02}
 \label{fig1}
 \end{center}
\end{figure}
%%%%%%%%%%%%%%%%%%%%%%%%   Figure  1  (end )  %%%%%%%%%%%%%%%%%%%%%%%%%

%%%%%%%%%%%%%%%%%%%%%%%%   Figure  2    %%%%%%%%%%%%%%%%%%%%%%%%%
%%%%%%%%%%%% two paths to type Ia SN explosion  and    DTD   %%%%%%%%%%%%%%%%%%%%%%%%

\begin{figure}[!tH]
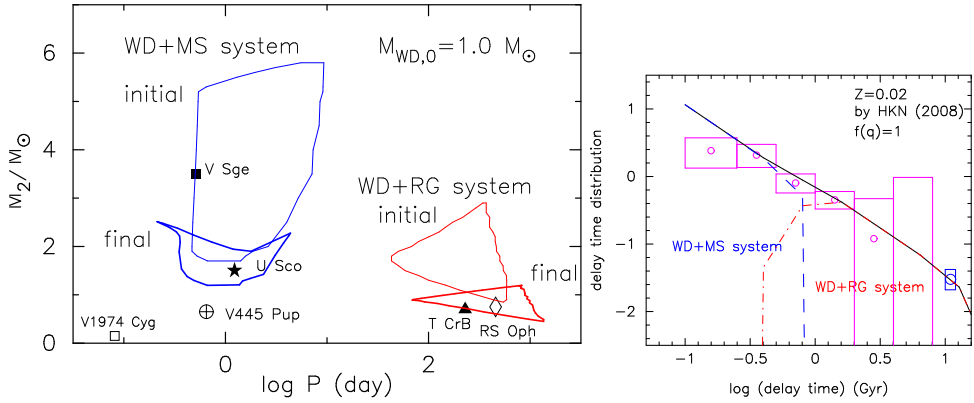

% \vspace*{-2.0 cm}
\begin{center}
 \includegraphics[width=3.in]{kato.fig2a.epsi} 
 \includegraphics[width=2.in]{kato.fig2b.epsi} 
% \vspace*{-1.0 cm}
 \caption{(left) The regions that produce SNe Ia are plotted in the 
orbital period -- secondary mass plane for the (WD + MS) system ({\it left}) 
and (WD + RG) system ({\it right}).
Here we assume the metallicity of $Z=0.02$ and
the initial WD mass of $M_{\rm WD, 0}= 1.0 ~M_\odot$.
The initial system inside the region encircled by a thin solid line
(labeled ``initial'') increases its WD mass up to the critical
mass ($M_{\rm Ia}= 1.38 M_\odot$) for the SN Ia
explosion, the regions of which are encircled by a thick solid line
(labeled ``final'').  Currently known positions of individual objects are 
plotted for recurrent novae:  
U Sco (\cite[Hacisu et al. 2000]{hkkm00}), T CrB (\cite{bel98}), RS Oph (\cite{bra09}), 
a supersoft X-ray source: V Sge (\cite[Hachisu et al. 2003]{hac03vsge}), classical novae: 
V1974 Cyg (\cite[Hachisu \& Kato 2006]{hac06a}), and  
the helium nova: V445 Pup (\cite{gor10,kat08v445}). 
(right) Delay-Time Distribution based on the SD scenario. 
Contributions of the two channels reproduce well the observational results 
(taken from \cite[Hachisu et al. 2008]{hac08}).
   \label{DTD}
}
\end{center}
\end{figure}

%%%%%%%%%%%%%%%%%%%%%%%%%%%%%%%%%%%%%%%%%%%%%%%%%%%

\firstsection % if your document starts with a section,
\section{Introduction}

Mass accreting white dwarfs (WDs) become novae, persistent 
supersoft X-ray sources (SSSs), or wind evolution objects, 
depending on the mass-accretion rate. 
Very massive WDs and its position in binary evolution 
are interesting subjects because 
they are closely related to Type Ia supernova (SN Ia) progenitors. 
In this paper I review how to specify the mass and its growth rate of WDs and 
discuss them in relation to progenitors of SN Ia. 
Section \ref{sec_evol} introduces binary evolution scenarios to SNe Ia in which 
accreting WDs are growing in mass (single degenerate scenario). 
For relatively low mass accretion rates ($\lesssim 1 \times 10^{-7}~M_\odot$ yr$^{-1}$), 
the WD experience recurrent nova outbursts.   
Section \ref{sec_rn} describes how to determine the WD mass from the light curves of  
novae and specifies mass increasing WDs.
With intermediate accretion rates hydrogen burning is stable and 
WDs become persistent supersoft X-ray sources (SSSs), 
which are the subject of Section \ref{sec_sss}.   
For high mass-accretion rates, optically thick winds inevitably occur.  
Such objects are described in Section \ref{sec_wind}. 
Section \ref{sec_henova} introduces helium novae and shows mass accumulation efficiency 
during helium shell-flashes.

\section{Binary Evolutions toward Massive WDs}\label{sec_evol}

Fig.\,\ref{fig1} shows the map of response of WDs to mass-accretion rates. 
When the mass-accretion rate onto the WD is smaller than $\dot M_{\rm st}$,  
hydrogen burning is unstable and nova outbursts (classical nova: CN) are triggered periodically. 
If the mass-accretion rate is relatively large and close to the upper limit, $\dot M_{\rm st}$, 
the WD experiences weak shell flashes with a short recurrence period. Such objects are 
recurrent novae (RNe). 
In the intermediate mass-accretion rate of 
$\dot M_{\rm st}~< \dot M_{\rm acc}~< \dot M_{\rm cr}$, 
hydrogen shell-burning is stable and the mass-accreting WD becomes a persistent SSS   
in which the mass-accretion rate is equal to the mass consuming rate owing to nuclear burning.
For a higher mass accretion rate ($\dot M_{\rm cr}~< \dot M_{\rm acc}$) 
hydrogen burning cannot consume all of the accreted material, the rest of which 
expands, and the optically thick winds blows. 
In traditional double degenerate (DD) scenario (e.g. \cite[Iben \& Tutukov 1984]{ibe84}), 
such a binary is considered to undergo the common envelope 
evolution that leads finally to a pair of degenerate stars. 
In the beginning of 1990, however, opacity tables are revised  
and the new opacities (e.g., OPAL opacities) show a strong peak at $\log T$ (K) $\sim 5.2$, 
which brings significant structure changes of WD envelopes, i.e.,  
the optically thick winds are accelerated and the envelope photosphere does 
not reach the companion orbit so the binary does not experience a common envelope evolution 
in $\dot M_{\rm cr}~<~\dot M_{\rm acc}~\lesssim 10^{-4}~M_\odot$ yr$^{-1}$. 
Therefore, the binary evolution was essentially changed in a large part of the upper region 
of Fig.\,\ref{fig1} (\cite[Hachisu et al. 1996]{hkn96}).

These optically thick winds were incorporated into binary evolution scenarios and  
two paths, in which the WD grows in mass 
toward the Chandrasekhar mass, were found. One is the WD + main sequence (MS) channel 
(\cite[Li \& van den Heuvel 1997; Hachisu et al. 1999a]{livan97,hknu99}) 
and the other is the WD + red giant (RG) channel  (\cite[Hachisu et al. 1999b]{hkn99}).

Fig.\,\ref{fig1} shows a typical evolutionary path to SN Ia explosion, 
which is common to both the channels. 
A binary starts the accretion wind phase, because 
the mass transfer rate onto the WD is 
larger than the critical rate for occurrence of optically thick 
winds ($\dot M_{\rm cr}$). So the binary undergoes the accretion wind evolution 
(see Section \ref{sec_wind} for more detail). 
As the mass transfer rate decreases with time, the wind stops and 
the binary enters the supersoft X-ray phase. Steady hydrogen burning occurs 
in both the wind and SSS phases, and the WD mass continuously increases.  
When the mass accretion rate further decreases to less than the value to keep steady
hydrogen-burning ($\dot M_{\rm st}$), the binary enters the recurrent nova
phase. The shell flashes are weak and a part of the accreted matter remains after 
each outburst, so the WD mass continuously increases. 
The WD explodes as an SN Ia at the star mark in Fig.\,\ref{fig1}.

Fig.\,\ref{DTD} (left) shows the initial/final regions of the binary evolutions 
in which the WD grows its mass to the critical mass ($M_{\rm Ia}= 1.38~M_\odot$) and
explodes as an SN Ia. The binary 
inside the regions (labeled ``initial'') evolves down to inside the ``final'' region and 
explode there. 

This figure also shows individual RNe, wind phase object (V Sge) 
and CN (V1974 Cyg), helium nova (V445 PUp) with the known orbital period 
and suggested secondary mass.  
The position of three RNe, U Sco, RS Oph and T CrB, are very consistent with 
the final region, and of V Sge (still midway of evolutional path downward) 
is also consistent. The classical nova V1974 Cyg locates below the final region 
of the WD+MS system, 
which indicates that they are not SN Ia progenitor (see Section \ref{sec_rn}). 

These two different paths to SNe Ia explain well observed delay time distribution 
as shown in Fig.\,\ref{fig1} (right). The WD+MS channel corresponds to the prompt component 
and WD+RG channel the tardy component.
(see \cite{hac08} for details). 
 
%%%%%%%%%%%%%%%%%%%%%%%%   Figure  3 and 4  (RS Oph and  V2491 Cyg) %%%%%%%%%%%%%%%%
\begin{figure}[t]
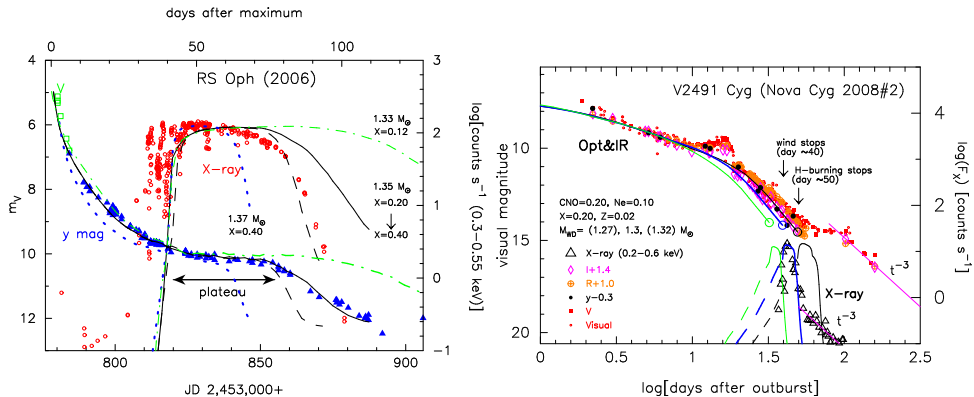

%\begin{figure}[b]
% \vspace*{-2.0 cm}
%\begin{center}
 \includegraphics[width=2.5in]{kato.fig3a.epsi}
 \includegraphics[width=2.5in]{kato.fig3b.epsi}
% \vspace*{-1.0 cm}
 \caption{(left)
Light curve fitting of the recurrent nova RS Oph. 
Observed light curve: 
Optical magnitude (green open squares and blue filled triangles), and   
supersoft X-ray count rates (red open circles). 
Lines indicate model light curves: $M_{\rm WD}= 1.33~M_\odot$ (dash-dotted line),
$1.35~M_\odot$ (solid line), and $1.37~M_\odot$ (dashed line). 
This figure is taken from \cite{hac07}.
(right) 
Light-curve fitting for the classical nova V2491 Cyg. 
The upper bunch of data indicates optical and near-IR observational data, and 
the lower black triangles X-ray data. 
The best-fit theoretical model is a $1.3~\mathrm{M_\odot}$ (thick blue line)
for the envelope chemical composition with 
$X=0.20$, $Y=0.48$, $X_{\rm CNO} =0.20$, $X_{\rm Ne} =0.10$, and $Z=0.02$.
Supersoft X-rays may not detected during the wind phase (dashed part) 
because of self-absorption by the wind itself.
The $F_\lambda \propto t^{-3}$ law is added for the
nebular phase. See \cite{hac09a} for more detail.
 \label{rsoph}
}
% \end{center}
\end{figure}
%%%%%%%%%%%%%%%%%%%%%%%%   Figure  3 and 4 (end )  %%%%%%%%%%%%%%%%%%%%%%%%%

%%%%%%%%%%%%%%%%%%%%%%%%%%%%%%%%%%%%%%%%%%%%%
\section{Recurrent novae}\label{sec_rn}

Nova light curves are calculated using the optically thick 
wind theory of nova outburst for a given set of the WD mass and chemical 
composition of the envelope  (\cite[Kato \& Hachisu 1994]{kat94h}). 
The optical and infrared (IR) fluxes are basically well represented by free-free 
emission. The super soft X-ray light curve is calculated from blackbody emission, 
which may be inaccurate but enough for the purpose to estimate the X-ray turn on/off time. 

Fig.\,\ref{rsoph} shows observed multi-wavelength light curve of RS Oph and V2491 Cyg. 
After the optical maximum, the photospheric temperature rises with time 
while the total luminosity is almost constant. Therefore, the main-emitting region 
of photon shifts from optical to UV and then supersoft X-ray. 
The decline rate of optical flux and durations of X-ray phase depend 
differently on the WD mass and composition. 
This property is useful to determine the WD mass. 

Fig.\,\ref{rsoph} (left) also shows the light-curve fitting of the recurrent nova RS Oph. 
The light curve model consists of a $1.35~M_\odot$ WD, a $0.7~M_\odot$ RG companion 
with a radius of $35~R_\odot$ and  
an irradiated disk with a radius of $47 ~R_\odot$
around the white dwarf. The binary orbital period is 455.72 days and  
the inclination angle is $i=33^\circ$.
The model produces a reasonable agreement with the observation of 
both the supersoft X-ray and optical observations (see 
\cite[Hachisu et al. 2006]{hac06} for more detail).  

The light-curve fitting of V2491 Cyg is shown in Fig.\,\ref{rsoph} (right). 
V2491 Cyg is a very fast CN that shows very similar early decline as in RS Oph 
(it may not look like because the left figure is plotted in the linear scale while 
the right one is in the logarithmic scale), but shows a very short   
supersoft X-ray phase of only 10 days. 
The best fit model, a $1.3~M_\odot$ WD, reproduces simultaneously   
the light curves of visual, IR, and X-ray, except the secondary  
maximum about 15 days after the optical peak, which was explained as 
a magnetic origin (see Hachisu and Kato 2009a). 
In the very later phase the visual light curve deviates from the theoretical lines due to 
contribution of strong emission lines.

In general, a nova evolves faster in a more massive WD and slower in a less massive WD. 
Therefore, a very fast decline of optical light curve is an indication of 
a very massive WD, but not all of the massive WDs are candidates of SN Ia 
 progenitor. 
For example, V2491 Cyg shows a fast optical decline in the early phase 
very similar to RS Oph, and the WD mass is estimated to be as massive 
as $\sim 1.35~M_\odot$ in RS Oph and $\sim 1.3~M_\odot$ in V2491 Cyg. 
RS Oph shows no indication of heavy element enhancement in its ejecta, which is consistent 
with the mass increasing WD. Detailed light curve analysis of RS Oph derived the mass increasing 
rate of WD  
is $(0.6-1) \times 10^{-7}~M_\odot$ yr$^{-1}$ (\cite[Hachisu et al. 2007]{hac07}). 
On the other hand,   
V2491 Cyg is a classical nova, and heavy element enhancement of ejecta 
is observed (\cite[Munari et al. 2011]{mun11}). This indicates that the WD mass is decreasing, 
because the surface layer of the WD was eroded during the outburst. 
 
It is interesting to compare the durations of SSS phase in RS Oph  
and V2491 Cyg. The SSS phase of RS Oph lasts as long as 60 days, whereas 
only 10 days in V2491 Cyg. The long lasted SSS phase in RS Oph is  
an indication of a hot helium ash layer produced by hydrogen burning. 
This hot helium ash act as a heat reservoir to keep the WD temperature high enough to 
emit supersoft X-rays for a long time (\cite[Hachisu et al. 2007]{hac07}). 
RS Oph shows a plateau phase in the optical light curve which is explained as 
a contribution of the irradiated disk (\cite{hac06}) heated by the hot WD. 
The sharp end of the plateau phase corresponds to the turnoff 
of supersoft X-ray (\cite[Hachisu et al. 2007]{hac07}).
V2491 Cyg shows a very short duration, which indicates absence of helium ash layer, 
consisting with the mass decreasing WD.

The classical nova V838 Her  also shows a very similar optical 
light curve to that of U Sco. V838 Her is a classical nova with very massive 
WD ($\sim 1.35~M_\odot$) and heavy element enrichment is observed in its 
ejecta (see Table 2 in \cite[Kato et al. 2009]{kat09}) This indicates that   
the WD mass is decreasing.
On the other hand,  U Sco is a recurrent nova in which the WD 
mass is increasing at an average rate of $\sim 1 \times 10^{-7}~M_\odot$ yr$^{-1}$  
(\cite[Hachisu et al. 2000]{hac00}). 
U Sco also shows a plateau phase of 18 days corresponding to the SSS phase. 
This is also the indication of helium heat reservoir, i.e., the WD mass is increasing.

It should be noticed that classical nova binaries undergo different evolutional paths  
from recurrent novae, and do not produce SN Ia. 
Fig.\,\ref{DTD} (left) shows that some classical novae are in the outside of the 
initial/final region of promising binary regions to SNe Ia, whereas 
RNe locate within the final region because their WD mass is close to the 
final value of $1.38~M_\odot$.

\section{Supersoft X-ray sources} \label{sec_sss}

In the intermediate mass-accretion rate
($\dot M_{\rm st} \le \dot M_{\rm WD} \le \dot M_{\rm cr}$), 
hydrogen burning is stable and optically thick winds do not occur.
\cite[van den Heuvel et al. (1992)]{van92} interpolated supersoft X-ray sources 
like Cal 83 and Cal 87 in Large Magellanic Cloud (LMC) as 
an accreting WD with high mass-accretion rate ($\approx\,10^{-7}\,M_\odot~$yr$^{-1}$) 
with steady hydrogen nuclear burning.  
Distribution of LMC SSS in HR diagram is consistent with theoretical SSS region 
of accreting WDs. The photospheric temperature of the WD is relatively high,  
$\log T$ (K)=5.4-6.0, so the WDs emit supersoft X-rays (\cite[Nomoto et al. 2007]{nom07}). 
Note that LMC SSSs are all close binaries, which correspond to the SSS phase of the 
WD + MS channel. 
There are two SSS symbiotic stars in SMC (SMC3 and Lin358). 
They may correspond to the SSS phase of the WD + RG channel, although the WD mass 
is not well determined. Their surface temperature is relatively low, $\log T$ (K)=5.5 
and 5.7 for Lin358 and SMC3, respectively, which may be one of the reason of 
relatively low X-ray emission.

Recently, supersoft X-ray fluxes of early type galaxies are obtained by 
\cite{GB10}. In early type galaxies the star formation ended long time ago, so the 
progenitor binaries are all the WD+RG systems. 
The X-ray fluxes estimated by \cite{GB10} are very consistent with model  
X-ray fluxes based on the SD scenario, which is a strong support to the SD scenario 
of SN Ia, although Gilfanov \& Bogd\'an  drew 
an opposite conclusion (see \cite{kat11sss,hac10} for more detail).

%%%%%%%%%%%%%%%%%%%%%%%%   Figure  4   (RX J0513)  %%%%%%%%%%%%%%%%%%%%%%%%%
\begin{figure}[b]
% \vspace*{-2.0 cm}
%\begin{center}
 \includegraphics[width=2.in]{kato.fig4a.epsi}
 \includegraphics[width=3.in]{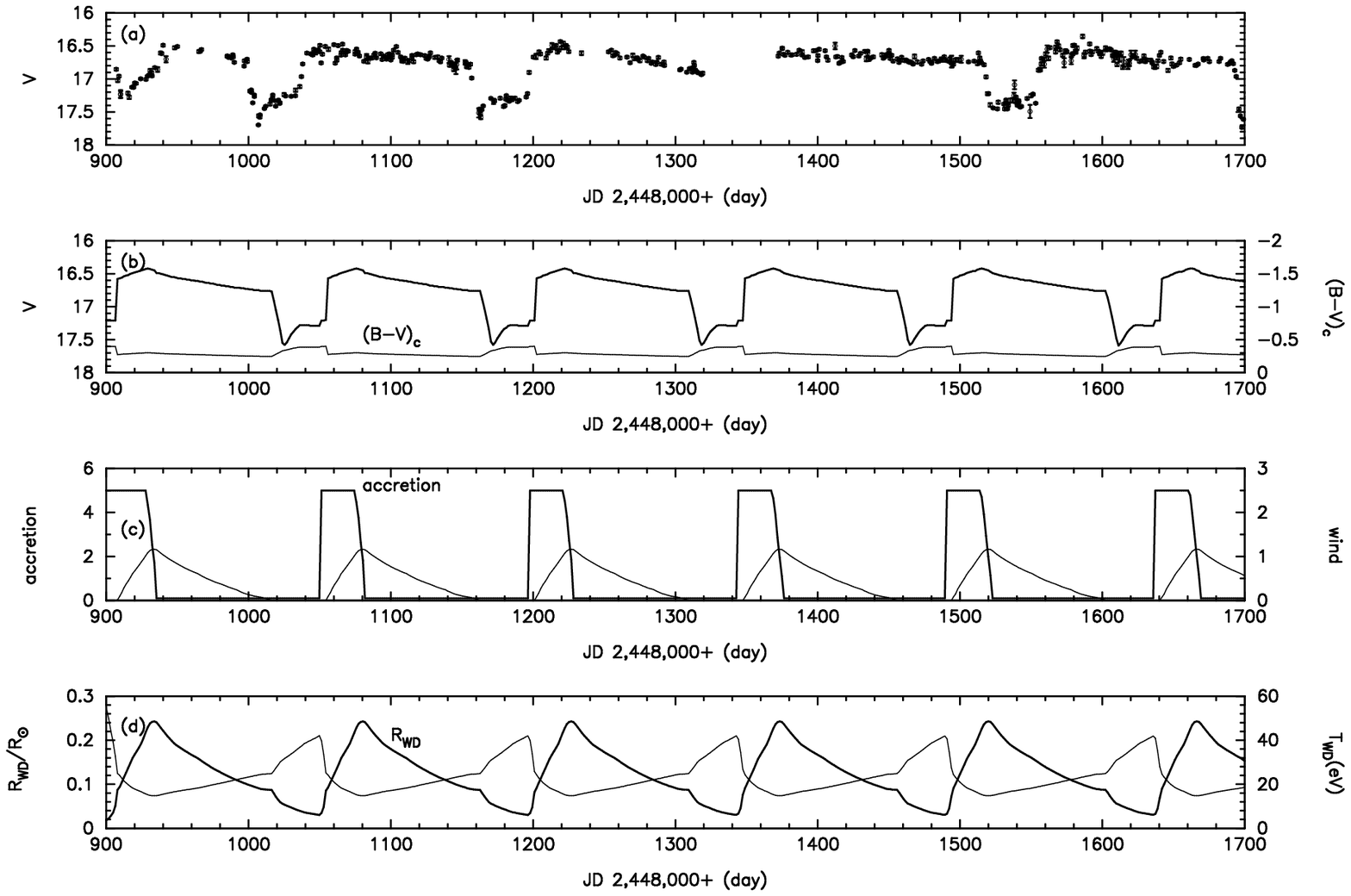}
% \vspace*{-1.0 cm}
 \caption{(left) 
Optically thick winds blow from mass-accreting  WDs 
when the mass-transfer rate from a lobe-filling companion exceeds a critical 
rate, i.e., $\dot M_{\rm acc} > \dot M_{\rm cr}$.  
The white dwarf accretes mass from the equatorial region and 
at the same time blows winds from the polar regions.
Stable hydrogen nuclear burning occurs on the surface of the WD. 
(right)Self-sustained model of spontaneous winds for supersoft X-ray source 
RX J0513-6951. (a) long term evolution of V magnitude. (b) model light curve of  
$M_{\rm WD}=1.3 M_\odot$. (c)change of accretion rate and wind mass-loss from WD envelope.
(d) Change of WD radius and its temperature. (Taken from \cite{hac03rxj}.)
  \label{rxj0513}
}
%\end{center}
\end{figure}

%%%%%%%%%%%%%%%%%%%%%%%%%%%%%%%%%%%%%%%%%%%%%

\section{Accretion wind phase} \label{sec_wind}

When the accretion rate is larger than $\dot M_{\rm cr}$, 
the WD cannot consume all of the accreted matter by nuclear burning, 
so the accreted matter is piled up to form an extended envelope. As the 
photospheric temperature decreases to below the critical value,  
 optically thick winds are accelerated due to 
Fe peak (at around $\log T ($K$)\approx 5.2$) of the 
OPAL opacity (Kato and Hachisu 1994, 2009).

Hachisu and Kato (2001) proposed a binary system in which the WD accretes 
matter from the companion from the equatorial region and loses matter as 
a wind from the other regions as illustrated in Fig.\,\ref{rxj0513} (left).
They called such a configuration ``accretion wind.'' 
In this situation, the WD burns hydrogen at the rate of steady burning $\dot M_{\rm nuc}$ and 
blows the rest of the accreted matter in the winds at the rate of about 
$\dot M_{\rm WD} - \dot M_{\rm nuc}$. 
WDs in this ``accretion wind''  
correspond to the upper part of Fig.\,\ref{fig1}. 

Two SSSs, RX\,J0513.9$-$6951 and V\, Sge, correspond to 
this accretion wind evolution. 
RX J0513.9$-$6951 is an LMC supersoft X-ray source that shows quasi-regular transition between 
optical high and low states as shown in Figure \ref{rxj0513} (right). 
Supersoft X-rays are detected only in the optical low states (Reinsch et al. 2000; 
Schaeidt, Hasinger, and Truemper 1993).
 
Hachisu and Kato (2003b) presented a transition mechanism between the high 
and low states. In the optical high state, the accretion rate is high enough  
and the photosphere expands to accelerate winds. The WD has a low surface  
temperature and no X-rays are expected. 
In the optical low state, mass-accretion rate is low and 
the photospheric temperature is high enough to emit supersoft X-rays. 
No wind is accelerated. 
The authors proposed a self-regulation transition 
mechanism that makes the binary 
back and forth between the optical high and low states.
When the mass-accretion rate is large, the WD is in the optical 
high state. The strong winds hit the companion and strip off 
a part of the companion surface. Thus the mass-transfer rate 
onto the WD reduces and finally stops, which causes the wind stop and 
the system goes into the optical low state. After a certain time, the 
companion recovers to fill the Roche lobe again and the mass transfer resumes, 
which cases wind mass loss. 
The resultant theoretical light curves depend on the WD mass and other 
parameters. The best fit model that reproduces 
the observed light curve indicates the WD 
mass to be 1.2 - 1.3 $\mathrm{M_\odot}$ (\cite[Hachisu \& Kato 2003a]{hac03rxj}).

The second object is V Sge that also shows the similar semi-regular transition 
of light curve, although timescales are different. 
Its light curve is also reproduced by the transition model with 
the WD mass of 1.2 - 1.3 $\mathrm{M_\odot}$ (\cite[Hachisu \& Kato 2003b]{hac03vsge}).

In these two systems the WD mass is increasing with time, because steady 
nuclear burning produces helium ash which accumulate on the WD. Therefore,   
they are candidates of SN Ia progenitors.
Fig.\,\ref{DTD} shows the position of V Sge in the initial/final binary diagram. 
The WD, of mass 1.2 - 1.3 $\mathrm{M_\odot}$, is still increasing and the position of 
V Sge is in the way of downward toward the final region. 

These two object, RX J0513.9-6951 and V Sge suggest that the accretion wind evolution and 
stripping effects of companion surface that regulate the mass-transfer rate from the 
companion are important and are actually working in binary evolution. 
This accretion wind is an important elementary process for binary evolution 
scenario to Type Ia supernova, because it governs the growth rate of the WD mass 
(e.g., \cite{hkn99,hknu99}, Han \&
Podsiadlowski 2006),  
as well as the mass-transfer rate from the companion which is regulated 
by stripping of companion surface by the wind 
(e.g., Hachisu, Kato \& Nomoto 2008). 

%%%%%%%%%%%%%%%%%%%%%%%%%%%%%%%%%%%%%%%%%%%%%
%%%%%%%%%%%%%%%%%%%%%%%%%%%%%%%%%%%%%%%%%%%%%
\begin{figure}[b]
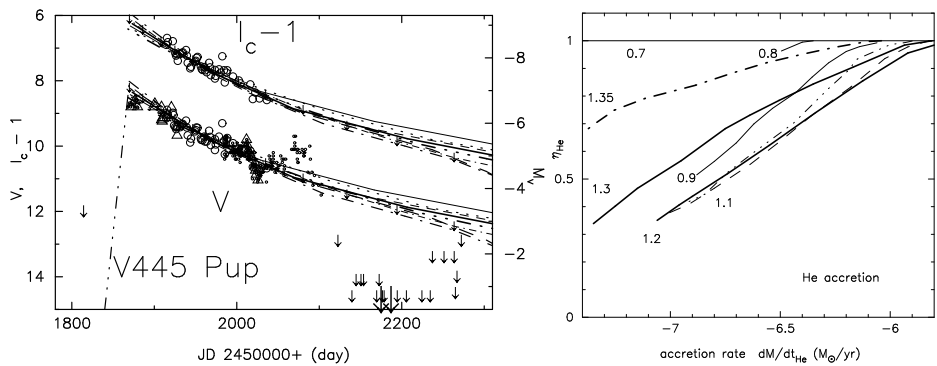

% \vspace*{-2.0 cm}
\begin{center}
 \includegraphics[width=2.8in]{kato.fig5a.epsi} 
 \includegraphics[width=2.in]{kato.fig5b.epsi} 
% \vspace*{-1.0 cm}
 \caption{(left) Light curve  fitting of the helium nova V445 Pup.
Lines indicate theoretical model for 1.35 $M_\odot$ (dotted line),  
1.377 $M_\odot$ (dashed and dash-dotted lines), and  
1.37 $M_\odot$ (other lines)
See \cite{kat08v445} for detail.
(right) Mass accumulation efficiency, $\eta_{\rm He}$, is plotted
against helium mass accretion rate. The WD mass is attached beside the curve.
Figure taken from \cite{kat04}.   \label{V445pup}
} 
\end{center}
\end{figure}

%%%%%%%%%%%%%%%%%%%%%%%%%%%%%%%%%%%%%%%%%%%%%

\section{He novae} \label{sec_henova}

Helium nova is a similar phenomenon to ordinary nova, but nuclear fuel is helium. 
If the companion is a helium star, the WD accretes helium and 
when the accreted helium mass reaches to a critical value, unstable helium 
burning triggers a thermonuclear runaway event. 
If the companion is a normal star, and the hydrogen accreting WD 
is growing in mass (like recurrent novae or SSS), a helium ash layer develops  
underneath the hydrogen burning zone, which also triggers a helium nova outburst. 
Such helium novae were theoretically predicted by \cite{kat89} long before 
V445 Pup was discovered in 2000 December, that has been the only helium nova so far.  

Fig.\,\ref{V445pup} (left) shows the light curve of V445 Pup, that shows a slow decline 
in $V$ and $I$ bands before the blackout due to dust formation.  We have only 
210 days data available for light curve analysis. 
From the light curve fitting the WD mass is estimated to be very massive 
($\geq 1.35 M_{\odot}$) and 
the WD is growing in mass (see \cite{kat08v445} for more detail). 
Therefore, V445 Pup is a candidate of SN Ia progenitors. 
The companion is an evolved helium star but its evolutional path is not well known; 
such a system of massive WD + helium star companion has no corresponding stage in 
the two known channels to SNe Ia (WD+MS channel and WD + RG channel).

Fig.\,\ref{V445pup} (right) shows the mass accumulation efficiency
$\eta_{\rm He}$, the ratio
of the processed matter remaining after one cycle
of helium shell flash to the ignition mass (\cite[Kato \& Hachisu 2004]{kat04}). 
In low accretion rates ($\log \dot M_{\rm He}(M_{\odot}$ yr$^{-1}) < -7.6$) a helium
detonation occurs which may cause a supernova explosion.
In steady state accretion, i.e., in the wind phase or SSS phase, the helium accretion 
rate is about -6.0 - -6.3 for $\gtrsim 1.3~M_\odot$ WDs, and  $\eta_{\rm He}$ is as 
high as $> 0.9$; most of the accreted matter accumulates on the WDs and 
the rest is lost in the wind. 
These mass accumulation efficiencies have been 
incorporated in many binary evolution calculations. 
(see also Discussion after the talk.)

\acknowledgments
This research has been supported by 
the Grant-in-Aid for Scientific Research of the Japan Society for the
Promotion of Science (22540254). 

% todayreference

\begin{discussion}

\discuss{Webbink}{During He shell flashes, a larger part of the 
envelope will be lost, which prevent WDs to grow to the Chandrasekhar mass.}

\discuss{Kato}{Your argument is based on the paper of Cassisi et al. (1998) which 
claim that WDs cannot attain the Chandrasekhar mass because most of the envelope is lost 
during H/He shell flashes due to the interaction of the expanded envelope with the companion star. 
I have counterarguments and comments on the paper as follows; \\
1. This Cassisi et al.'s argument is just an assumption. They did not include frictional 
process in their calculation but simply assumed that the most part of the envelope may be 
lost, once the envelope expands beyond the companion orbit. 
However, Kato had already shown that frictional process is ineffective in shell flashes 
before their work (Kato 1991, ApJ 373, 620:Kato 1991, ApJ 383,761).\\
2. Nova light curves show no indication of frictional process. 
If the companion motion is so effective in mass ejection of the envelope, there should be 
a clear indication associated with the epoch when the companion emerges from the extended 
envelope. 
However, there are no such effects in H/He nova outbursts (Kato \& Hachisu 2011, 
submitted to ApJ). \\
3. Cassisi et al. used the old opacity. It is strange that they emphasized that 
the Los Alamos opacities are very similar to the OPAL opacities, whereas 
OPAL opacities have a famous large peak at $\log T ({\rm K}) \sim 5.2$, while 
Los Alamos opacities do not.   
If one use the OPAL opacity, the optically thick winds is accelerated on which frictional effects 
are very small due to small density at the orbit (see Kato \& Hachisu 1994).  \\
4. Cassisi et al. adopted 0.5 and 0.8 $M_\odot$ WDs and claimed that these 
are typical initial masses for mass increasing WDs toward the Chandrasekhar mass. 
However, modern SD scenarios (e.g., Hachisu et al., 1999, ApJ, 519, 314: Hachisu et al., 
1999, ApJ,522,487) show that a possible range of the WD mass is more massive 
(namely, 0.9-1.0 $M_\odot$), and a low mass WDs such as 0.5 -0.8 $M_\odot$ 
are unlikely to be the candidate of SN Ia.

From these reasons, I feel Cassisi' et al.' (1998)'s claim is very strange. 
Nevertheless, this Cassisi et al.'s arguments, have been still cited  
as a counterargument to the SD scenario, in order to to draw a conclusion 
that accreting WDs cannot reach the Chandrasekhar mass
because of mass ejection due to frictional process during shell flashes.  
% \cite[e.g. {yun00,yun05,bad07,mao10}.
I found, unfortunately, other presentations of this conference 
(e.g., Nelemans, Yungelson) adopt this Cassisi et al.'s argument 
as a counterargument to the SD scenario, but this is illogical.
}
\end{discussion}

\end{document}